\documentstyle[aps,prl,epsfig]{revtex}

\newcommand{\Dslash}{{D \hskip -6pt /}}

\bibstyle{unsrt}
\begin{document}
\draft

\title{Chiral Extrapolation: An Analogy with Effective Field Theory}
\author{Gerald V. Dunne}
\address{Department of Physics, University of Connecticut, Storrs CT
06269, USA}
\author{Anthony W. Thomas and Stewart V. Wright}
\address{Special Research Centre for the Subatomic Structure of
Matter, \\ and Department of Physics and Mathematical Physics, \\
University of Adelaide, Adelaide SA 5005, Australia}

\maketitle
\vskip .5cm

\begin{abstract} We draw an analogy between the chiral
extrapolation of lattice QCD calculations from large to small quark masses
and the interpolation between the large mass (weak field) and small mass
(strong field) limits of the Euler--Heisenberg QED effective action. In
the latter case, where the exact answer is known, a simple extrapolation
of a form analogous to those proposed for the QCD applications is shown to
be surprisingly accurate over the entire parameter range. 
\end{abstract}

\section{Introduction}

The challenge to find an accurate and reliable method of chiral
extrapolation for hadronic properties calculated in lattice QCD at large
quark mass is a matter of considerable current importance. While
computer limitations mean that lattice simulations
at physical quark masses are many years away, recent progress in
chiral extrapolation suggests that it may well
be possible to obtain accurate hadronic properties based on the
calculations which will be possible with the next generation of
supercomputers, available within just a few years, in the 10 Tera-flops range.

{}Fundamental to this scheme is the development of 
extrapolation methods which incorporate the model independent constraints
of chiral symmetry \cite{extrap,extrap2}, notably the leading non-analytic
(LNA) behaviour of chiral perturbation theory \cite{chext,chipt}, as well as the
heavy quark limit \cite{shifman}. 
Although these extrapolations are designed to match the leading behaviour
in the extreme limits of small and large quark mass, there has been
little guidance as to their reliability in the intermediate mass region.
It is very unclear what precision to expect from
such a simple extrapolation into the intermediate mass region, because the large
mass expansion is presumably asymptotic, and the small mass limit has a log
divergence plus finite corrections 
with a small radius of convergence. Here we attack
this question from a novel direction by considering a remarkably close analogy
between this problem and a well-known, exactly soluble system in effective field
theory --- the Euler--Heisenberg effective action
\cite{euler,weisskopf,schwinger}. The Euler--Heisenberg system exhibits many of
the features found in the QCD calculations: at small electron mass
(equivalently, strong external field) there is a logarithmic branch point, while
at large mass (equivalently, weak external field) one has an asymptotic series
expansion in inverse powers of mass. In this Letter, we show that a simple
two-parameter interpolation formula (of the form used in the 
context of chiral extrapolation),
which builds in the correct leading behaviour in both the small and large mass
limits, yields an excellent approximation to the exact Euler--Heisenberg answer
over the entire range of mass. We discuss possible consequences of this
observation for the chiral extrapolation of lattice data.

Effective field theory (EFT) plays an important role in modern
theoretical physics \cite{weinberg,donoghue,manohar}. In pioneering work in the
1930's, Heisenberg and Euler \cite{euler}, and Weisskopf \cite{weisskopf},
studied the quantum corrections to classical electrodynamics 
associated with vacuum
polarization effects. Renormalization properties and a more formal
``proper-time'' version were later studied by 
Schwinger \cite{schwinger}.
In modern language, they computed the low energy effective action for the
electromagnetic field, to leading order in the derivative expansion, by
integrating out the electron degrees of freedom in the presence of a
constant background electromagnetic field. This one-loop effective action
can be expressed as \cite{dittrich}
\begin{equation} 
S=-i \ln \det (i\Dslash-m) ,
\label{action}
\end{equation} 
where $\Dslash =\gamma^\nu\left(\partial_\nu+ie
A_\nu\right)$, and $A_\nu$ is the fixed classical gauge potential with
field strength tensor $F_{\mu\nu}=\partial_\mu A_\nu- \partial_\nu A_\mu$.
As shown in \cite{euler,weisskopf,schwinger}, this effective action can be
computed in a simple closed form when the background field strength
$F_{\mu\nu}$ is constant. For simplicity, we consider the case when the
background is a constant magnetic field of strength $B$ (and we choose
$eB$ to be positive). Then the exact, renormalized, one-fermion-loop effective
action has the following integral representation:
\begin{equation} 
S=-\frac{e^2 B^2}{8\pi^{2}} \int_{0}^{\infty}
\frac{ds}{s^{2}}\;  \left(\coth s-\frac{1}{s}-\frac{s}{3}\right)
\,e^{-m^2s/(eB)} . 
\label{proper}
\end{equation} 
The $\frac{1}{s}$ term is a subtraction of the zero field ($B=0$)
effective action, while the $\frac{s}{3}$ subtraction corresponds to a
logarithmically divergent charge renormalization \cite{schwinger}. 

We stress 
that Eq.\,(\ref{proper}) is an exact, non-perturbative result. However, it can
of course be expanded in two obvious limits. In the large mass limit,
$m^2\gg eB$ (which is equivalently the weak field limit), it is
straightforward to develop an (asymptotic) expansion of this integral:
\begin{eqnarray} 
S&=& -\frac{2 e^2 B^2}{\pi^2}\left(\frac{e
B}{m^2}\right)^2 \sum_{n=0}^\infty 
\frac{2^{2n}{\cal B}_{2n+4}}
{(2n+4)(2n+3)(2n+2)}\left(\frac{eB}{m^2}\right)^{2n}
\nonumber\\ &=&-\frac{e^2 B^2}{8\pi^{2}}\left[ -{1\over
45}\left(\frac{eB}{m^2}\right)^2 + {4\over 315}
\left(\frac{eB}{m^2}\right)^4 - {8\over 315}
\left(\frac{eB}{m^2}\right)^6 + \dots \right] 
\label{eh}
\end{eqnarray} 
Here the ${\cal B}_{2n}$ are the Bernoulli numbers \cite{gradshteyn}. The
large mass expansion, Eq.\,(\ref{eh}), of the effective action has the standard
form, $S=m^4 \sum_n \frac{{\cal O}^{(n)}}{m^n}$, of a low energy effective
action \cite{weinberg,donoghue}, where the higher-dimension operators
${\cal O}^{(n)}$ (of dimension $n$) are balanced by $n$ powers of the mass
scale $m$, below which the low energy effective action is meaningful. In
this case, ``low energy'' means that the cyclotron energy scale
$\frac{eB}{m}$ is much smaller than the energy scale 
set by the electron mass $m$. That is, $\frac{eB}{m^2}\ll 1$. An alternate
perspective on the large mass expansion is as a perturbative expansion in
powers of the coupling
$e$, with the $n^{\rm th}$ power of $e$ being associated with a
one-fermion-loop diagram with $n$ external photon lines (the divergent
${\rm O}(e^2B^2)$ self-energy term is not included, as it contributes to
the bare action by charge renormalization \cite{schwinger}). We note that,
as a consequence of charge conjugation (Furry's theorem), only even
powers of $\frac{eB}{m^2}$ appear in the perturbative expansion of
Eq.\,(\ref{eh}).  It is interesting to note that the series expansion of
Eq.\,(\ref{eh}) is divergent, because the Bernoulli numbers grow
factorially as
${\cal B}_{2n}\sim 2(-1)^{n+1} \frac{(2n)!}{(2\pi)^{2n}}$ for large $n$, consistent with
very general results for
perturbation theory \cite{dyson,zinn}. It is in fact an asymptotic series, 
and the proper-time integral representation in Eq.\,(\ref{proper}) is just the
straightforward Borel sum \cite{carl} of this asymptotic series \cite{dh}. 

The large mass limit may equivalently be characterized by the relevant
length scales: the electron Compton wavelength
$\lambda_e=\frac{1}{m}$, and the cyclotron radius (``magnetic length'')
$\lambda_B=\frac{1}{\sqrt{eB}}$. In terms of these length scales, the large
mass limit corresponds to the situation where the electron Compton
wavelength is much smaller than the cyclotron
radius: $\lambda_e\ll\lambda_B$.

Since the Euler--Heisenberg system is exactly soluble, we can also use the
exact integral representation (\ref{proper}) to study the small mass, or
strong field, limit where $m^2\ll eB$. In terms of the length scales, in
this limit the electron  Compton wavelength is much greater than the
cyclotron radius: $\lambda_e\gg\lambda_B$. Then, from Eq.\,(\ref{proper}), one
finds (using results in Ref.~\cite{matt}):
\begin{eqnarray}
S&=& -\frac{e^2 B^2}{8\pi^{2}}
\left\{\left[\frac{1}{3}+\frac{m^2}{eB}+
\frac{1}{2}\left(\frac{m^2}{eB}\right)^2\right]\log\frac{m^2}{eB}
+ \left[\frac{1}{3}-\frac{1}{3}\log
2-4\zeta^\prime(-1)\right]+\left[\log
\pi-1\right]\frac{m^2}{eB}\right.\nonumber\\
&& \qquad\qquad\qquad\qquad\left.
+\left[-\frac{3}{4}+\frac{\gamma}{2}-\frac{1}{2}\log
2\right]\left(\frac{m^2}{eB}\right)^2 -4 \sum_{k=2}^\infty
\frac{(-1)^k \zeta(k)}{k(k+1)}\left(\frac{m^2}{2 e
B}\right)^{k+1}\right\}\nonumber\\\nonumber\\
&=&-\frac{e^2 B^2}{8\pi^{2}}\left\{\frac{1}{3}\log\frac{m^2}{eB}
+0.763969  + {\rm O}\left(\frac{m^2}{eB},\frac{m^2}{eB}\log
\frac{m^2}{eB}\right)\right\} . 
\label{strongfield}
\end{eqnarray}
Note that the coefficient, $-\frac{e^2 B^2}{24\pi^2}$, of the leading term,
the $\log\frac{m^2}{eB}$ term, is fixed by the (one-loop) QED beta function
\cite{ritus}. In (\ref{strongfield}), $\gamma$ is Euler's
constant, and $\zeta(s)$ is the Riemann zeta function \cite{gradshteyn}.
Note that $\zeta^\prime(-1)\approx -0.165421$. 

\begin{figure}[hbt]
{\par\centering
\resizebox*{0.7\textwidth}{!}{\rotatebox{90}{\includegraphics{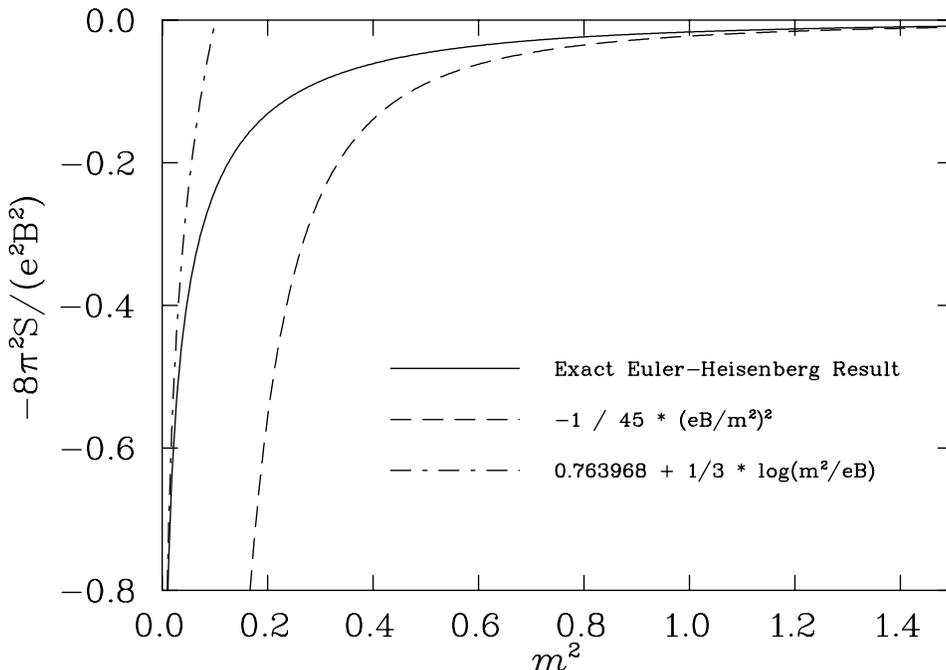}}} 
\par}
\caption{Comparison between the exact action (solid curve) for 
the Euler-Heisenberg
model and the leading terms in the expansions about the weak (dashed
curve) and strong field (dash-dot curve) limits. Note that $m^2$ is measured in
units of $eB$.
\label{fig:LeadingCurves}}
\end{figure}

It is instructive to contrast this small mass expansion,  
Eq.\,(\ref{strongfield}),
with the large mass expansion, Eq.\,(\ref{eh}). In the small mass limit,
analogous to the chiral limit in QCD, we see the appearance of 
logarithmic terms, analogous to the ``chiral logs'' of QCD. In addition, note
that both even and odd powers of $\frac{m^2}{eB}$ appear in the small mass
expansion, Eq.\,(\ref{strongfield}).
On the other hand, in the large mass
expansion, Eq.\,(\ref{eh}), there are no non-analytic log terms, and only even
powers of
$\frac{eB}{m^2}$ appear. So, we see that the 
one-loop Euler--Heisenberg effective
action, which is given by the exact integral 
representation (\ref{proper}), has
two very different expansions in the two limits 
of large and small electron mass.
The transition between these two extreme regions is governed by whether the
electron Compton wavelength, 
$\lambda_e$, is larger or smaller than the cyclotron radius, $\lambda_B$. In
{}Fig.~\ref{fig:LeadingCurves} we plot the exact Euler--Heisenberg
effective action,
Eq.\,(\ref{proper}), with an overall factor of $-\frac{e^2 B^2}{8\pi^2}$
removed, as a function of $\frac{m^2}{eB}$, and compare it to the leading
large mass term $-{1\over 45}\left(\frac{eB}{m^2}\right)^2$ from
Eq.~(\ref{eh}), and to the leading small mass terms
$\frac{1}{3}\log\frac{m^2}{eB} +0.763969$ from Eq.~(\ref{strongfield}).
{}From this figure it is clear that these leading terms accurately capture
the extreme behaviours of the exact result, but do not interpolate in the
intermediate region where the scales are comparable.

Having reviewed these pertinent aspects of the Euler--Heisenberg effective
action, we now turn to what appears at first glance to be a completely
different problem: the calculation of hadron properties as a function of
quark mass, or through the Gell-Mann--Oakes--Renner relation ($m_\pi^2
 \propto m_q$), pion mass. Chiral perturbation theory permits a rigorous
expansion of hadron properties about the chiral limit, where $m_\pi\to 0$.
For example, for the nucleon charge radius one finds \cite{Leinweber:1993hj}
\begin{equation}
\langle r^2\rangle_E = c_1 \pm \chi_N \, \log\frac{m_\pi}{\mu}
+c_2 m_\pi^2 +\dots
\label{chlimit}
\end{equation}
where $\pm$ refers to the proton or neutron respectively. (Here $\mu$
just sets the scale against which the pion mass is measured. It is
arbitrary in the sense that a change in $\mu$ is equivalent to a change
in the constant term, $c_1$.) Note that the charge radius diverges
logarithmically in the chiral limit, with a model independent coefficient
\begin{equation}
\chi_N=-\frac{(1+5g_A^2)}{(4\pi f_\pi)^2} . 
\label{chin}
\end{equation}
On the other hand, in the large $m_\pi$ limit, heavy quark effective
theory suggests that the charge radius should decrease as
\begin{equation}
\langle r^2\rangle_E=\frac{\bar{c}}{m_\pi^2}+\dots
\label{heavyq}
\end{equation}
%
plus higher inverse powers of $m_\pi^2$. (In the heavy quark limit one
has essentially a Coulombic problem and the charge radius is
proportional to the Bohr radius which goes as $1/m_q$ and hence 
$1/m_\pi$.)

As discussed at length in Ref.~\cite{chext}, current lattice data for
charge radii are confined to pion masses greater than 600 MeV. The
corresponding pion Compton wavelength, $\lambda_\pi$, is then smaller than
the calculated charge radius, which we may take as an indication of the
size, $R$, of the source of the pion field. The lattice data shows only a
very slow variation of $\langle r^2\rangle_E$ 
in the mass range where the lattice calculations have
been made, with no indication of a chiral log. Yet, in order to compare
with the physical charge radii one must extrapolate these lattice results
to the chiral regime where $\lambda_\pi\gg R$ and the chiral log is
important. This is the challenge of chiral extrapolation.

We wish to draw an analogy between the Euler--Heisenberg system
discussed above and this system. In this analogy, the pion Compton wavelength,
$\lambda_\pi$, plays the role of the electron Compton wavelength,
$\lambda_e$, and the source size, $R$, plays the role of the magnetic
cyclotron radius, $\lambda_B$, (equivalently, the mass scale $\mu^2$ plays
the role of the magnetic field strength $eB$). The chiral perturbation
theory expansion of Eq.\,(\ref{chlimit}), 
where $\lambda_\pi\gg R$, is analogous to
the leading terms in the small mass expansion of Eq.\,(\ref{strongfield}),
where
$\lambda_e\gg\lambda_B$. The heavy quark effective theory result
presented in Eq.\,(\ref{heavyq}), where $\lambda_\pi\ll R$, is
similarly analogous to the leading term in the large mass expansion in
Eq.\,(\ref{eh}) where $\lambda_e\ll\lambda_B$.

\begin{figure}[hbt]
{\par\centering
\resizebox*{0.7\textwidth}{!}{\rotatebox{90}
{\includegraphics{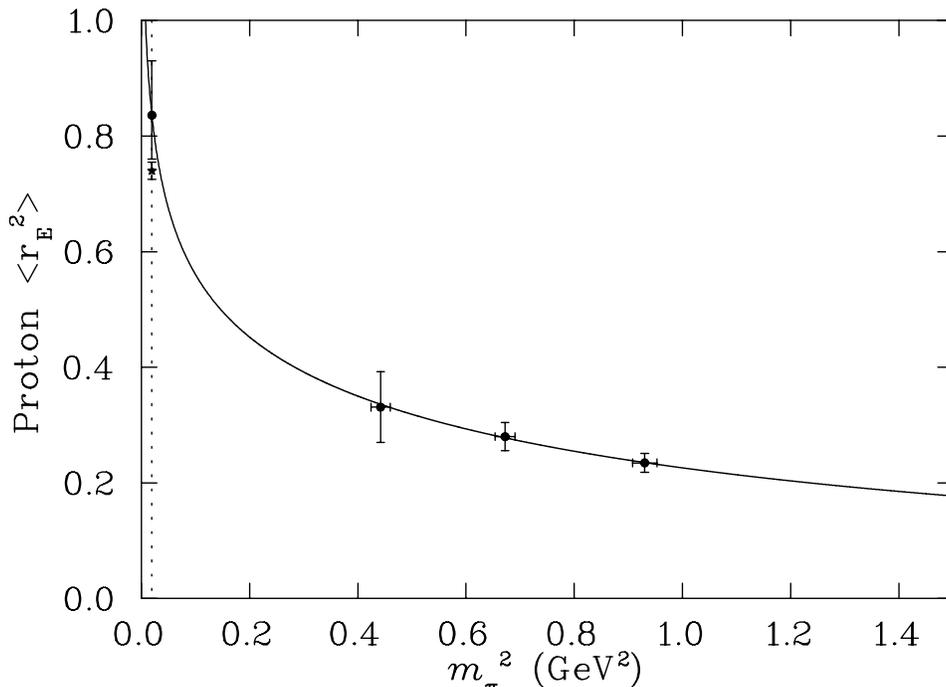}}} 
\par}
\caption{Fit to the lattice QCD data for the square of the 
proton charge radius as a function of pion mass squared, using
Eq.\,(\protect \ref{interp}). The extrapolated value at the physical pion mass
(indicated by the vertical dotted line) is shown by the solid dot with
the large error bar, while the star indicates the experimentally
observed value.
\label{fig:ChargeRadii}}
\end{figure}

In the QCD context, following earlier studies of magnetic 
moments \cite{extrap2}, where
it was found that a simple Pad\'e approximant was able to describe the
mass dependence arising in a particular chiral quark model, Hackett-Jones
{\it et al.} \cite{chext} extrapolated the lattice data from
$m_\pi^2>0.4\,\, GeV^2$ to $m_\pi^2=0.02\,\, {\rm GeV^2}$ (the physical point)
using an interpolating formula which was chosen as the simplest
two-parameter form consistent with the constraints imposed by the
extreme behaviours in the large and small pion mass limits,
Eq.\,(\ref{heavyq}) and Eq.\,(\ref{chlimit}) respectively.
(Recall that $\chi_N$ is model independent, and
note that the data could constrain no more than two parameters.)
In the light of later experience \cite{extrap}, we choose to use a
slightly modified argument in the chiral log:
\begin{equation}
\langle r^2\rangle_E={c_1\pm \frac{\chi_N}{2} \, 
\log\frac{m_\pi^2}{\mu^2 + m_\pi^2}\over
1+\bar{c}_2 m_\pi^2}.
\label{interp}
\end{equation}
Here, rather than being arbitrary, $\mu$ assumes physical
significance as the scale above which the chiral log is suppressed ---
of course, Eq.\,(\ref{interp}) preserves the correct behaviour in the
chiral limit. From experience with moments of structure functions,
magnetic moments and hadron masses, this scale is expected to be $\mu
\sim 500$ MeV. As the lattice data is not yet able to constrain $\mu$, we
simply fix it to 500 MeV and adjust only $c_1$ and $\bar{c}_2$. Figure 
\ref{fig:ChargeRadii} shows the resulting fit to the proton charge radius
and the corresponding extrapolation to the physical pion mass. As discussed in
\cite{chext}, this chiral extrapolation fit is closer to the physical value than
a naive linear fit through the lattice data. However, in the absence of
lattice data at lower quark masses, it is difficult to be more precise
about the quality of the fit.

In view of the close parallel between this hadronic problem and the
Euler--Heisenberg system in QED, we return to the Euler--Heisenberg system, where
we can be much more quantitative concerning the accuracy of an interpolating fit.
We ask the following question. Suppose that we did not know the exact integral
representation answer (\ref{proper}) for the effective action, but that we did
know the {\it leading} terms in each of the extreme large and small mass limits.
Would it then be possible to find a simple two-parameter interpolating formula,
analogous to (\ref{interp}), that connected the extreme limits in a smooth
manner? And if so, how accurate would such an interpolating formula be in the
intermediate region? 

\begin{figure}[hbt]
{\par\centering
\resizebox*{0.7\textwidth}{!}{\rotatebox{90}
{\includegraphics{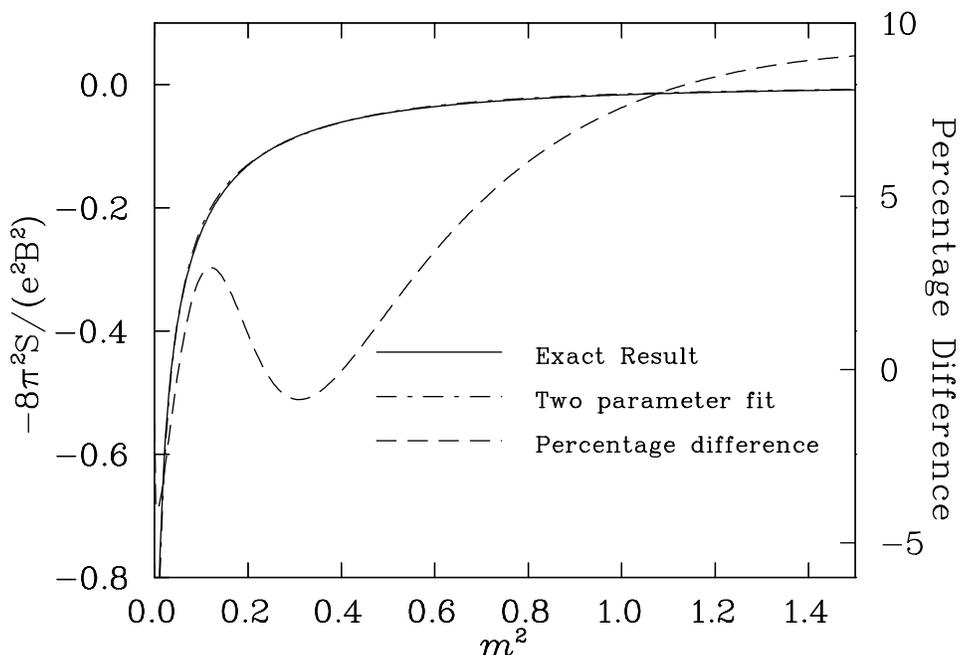}}} 
\par}
\caption{Comparison between the exact expression for the action in the
Euler-Heisenberg model (solid line) and the interpolating approximation
given in  Eq.\,(\protect \ref{interp2}) which builds in the correct
chiral and heavy quark limits (dot--dashed line). Note that the agreement
is so good that it is difficult to distinguish between the two curves on
this scale. The percentage difference between the two is indicated by the
dashed line.
\label{fig:Pade}}
\end{figure}

The leading terms are determined as follows. In the large mass limit, this
is the first term, $\frac{m^4}{360\pi^2}\left(\frac{eB}{m^2}\right)^4$, in
(\ref{eh}), corresponding to the first nonlinear correction to classical
electrodynamics, whose coefficient comes from the one-fermion loop
with four external photon lines, a straightforward perturbative
calculation. In the small mass limit, the leading term in
(\ref{strongfield}) is the logarithmic term,
$-\frac{m^4}{24\pi^2}\left(\frac{eB}{m^2}\right)^2
\log\frac{m^2}{eB}$, whose coefficient is fixed by the one-loop QED beta
function \cite{ritus}. Motivated by the 
interpolation formula, Eq.\,(\ref{interp}), which was 
used in the QCD case, we propose the
following interpolating function for the effective action
\begin{equation}
S_{\rm
interpolating}=-\frac{e^2B^2}{8\pi^2}\left({d_1+
\frac{1}{3}\log\left(\frac{m^2}{m^2+eB}\right)
- d_2\,\frac{m^2}{eB}\over 1+45 d_2 \left(\frac{m^2}{eB}\right)^3}
\right) .
\label{interp2}
\end{equation}
This interpolating formula has the correct leading behaviour in 
both the large and
small $m$ limits. Figure \ref{fig:Pade} shows a comparison of the fit obtained
with this form by adjusting the two parameters $d_1$ and $d_2$ (dash--dot curve)
with the exact result (solid curve). Our best fit was obtained with
parameter values: $d_1=0.7059$, and $d_2=1.5541$. Figure 3 also shows
the percentage difference between the exact result and approximate
expressions (dashed line). (Note that $m^2$ is expressed in units of
$eB$.) Over the entire range of
$\frac{m^2}{eB}$, the interpolating function is within 10\% of the
exact answer. Such precision is very surprising
when we recall that the Euler-Heisenberg effective action has the problems
(shared by the analogous QCD calculations) that the large mass expansion
is asymptotic and the small mass expansion has a log divergence and a
small radius of convergence.

In summary, the Euler-Heisenberg system presents a
problem which exhibits many of the mathematical complications of the 
chiral extrapolation problem in QCD, yet it is exactly soluble.
By carefully respecting both the high
and low mass limits of the exact solution, we showed how to construct a
simple formula which reproduced the exact solution over
the entire parameter range with surprisingly good accuracy.
Of course, in the Euler-Heisenberg case we have the advantage of fitting
the exact function over the entire mass range, while in QCD we have to
extrapolate from large quark mass (where lattice data is available) to
the chiral limit. Nevertheless, the fact that the
the mathematical structure of the two problems is identical, combined
with the success achieved in the Euler-Heisenberg problem, gives us
considerable 
confidence that a similar level of accuracy may be obtainable for QCD. 
It is therefore extremely encouraging that the chiral extrapolation of
even the present crude lattice data at very large quark masses yields a
physical proton charge radius within one standard deviation of the
experimental value. Even more important, this result
lends enormous impetus to the quest for new lattice data at lower
quark mass which will better constrain the chiral extrapolation. It
suggests that the next generation of supercomputers 
(available within 2--3 years) may well provide
sufficient information that, in combination with these chiral
extrapolation techniques, one should be able to calculate accurate
hadron properties at the physical quark mass. 

\vskip 1cm

\section{Acknowledgements:} GD is supported by the U.S.
Department of Energy grant DE-FG02-92ER40716.00, and thanks the CSSM at
Adelaide for hospitality and support while this work was begun.
This work was also supported by the Australian Research Council and the
University of Adelaide.


\begin{references}

\bibitem{extrap}

W.~Detmold, W.~Melnitchouk, J.~W.~Negele, D.~B.~Renner and A.~W.~Thomas,
Phys.\ Rev.\ Lett.\ {\bf 87}, 172001 (2001), [hep-lat/0103006];
D.~B.~Leinweber, A.~W.~Thomas, K.~Tsushima and S.~V.~Wright,
Phys.\ Rev.\ D {\bf 61}, 074502 (2000)
[hep-lat/9906027].
%
\bibitem{extrap2}
D.~B.~Leinweber, D.~H.~Lu and A.~W.~Thomas,
Phys.\ Rev.\ D {\bf 60}, 034014 (1999)
[hep-lat/9810005].
%
\bibitem{chext} 
E.~J.~Hackett-Jones, D.~B.~Leinweber and A.~W.~Thomas,
Phys.\ Lett.\ B {\bf 494}, 89 (2000)
[hep-lat/0008018].
%
\bibitem{chipt}
V.~Bernard, N.~Kaiser and U.~Meissner,
Int.\ J.\ Mod.\ Phys.\  {\bf E4}, 193 (1995)
[hep-ph/9501384].
%
\bibitem{shifman} M. Shifman, ``Recent Progress in Heavy Quark Theory'',
[hep-ph/9505289] lectures at TASI 95, published in {\it QCD and Beyond}, D. E.
Soper (Ed.), World Scientific, 1996.
%
\bibitem{euler} W. Heisenberg and H. Euler, ``Folgerungen aus der
Diracschen Theorie des Positrons'', Z. Phys. {\bf 98}, 714 (1936).
%
\bibitem{weisskopf} V. Weisskopf, ``Uber die Elektrodynamik des Vakuums
auf Grund der Quantentheorie des Elektrons'', Kong. Dans. Vid. Selsk.
Math-fys. Medd. XIV No. 6 (1936), reprinted in {\it Quantum
Electrodynamics}, J. Schwinger (Ed.) (Dover, New York, 1958).


\bibitem{schwinger} J.Schwinger, ``On Gauge Invariance and Vacuum
Polarization'', Phys. Rev. {\bf 82}, 664 (1951).

\bibitem{weinberg} S. Weinberg, {\it The Quantum Theory of Fields}, Vol.
II, (Cambridge University Press, Cambridge, 1996).

\bibitem{donoghue} J. Donoghue, E. Golowich and B. Holstein, {\it
Dynamics of the Standard Model} (Cambridge Univ. Press, 1992).

\bibitem{manohar} A. Manohar, ``Effective Field Theories'', Lectures at
the Schladming Winter School, March 1996 [hep-ph/9606222].

\bibitem{dittrich} W. Dittrich and M. Reuter, {\it Effective Lagrangians in
Quantum Electrodynamics}, Lecture Notes in Physics, Vol. 220, (Springer, Berlin,
1985).

\bibitem{gradshteyn} I. S. Gradshteyn and I. M. Ryzhik, {\it Table of
Integrals, Series, and Products}, Academic Press, New York, 1979.

\bibitem{dyson} F. J. Dyson, ``Divergence of Perturbation Theory in
Quantum Electrodynamics'', Phys. Rev. {\bf 85}, 631 (1952).

\bibitem{zinn} J. C. Le Guillou and J. Zinn-Justin (Eds.), {\it
Large-Order   Behaviour of Perturbation Theory}, (North Holland,
Amsterdam, 1990).

\bibitem{carl} C. M. Bender and S. A. Orszag, {\it Advanced Mathematical
Methods for Scientists and Engineers} (McGraw-Hill, New York, 1978).

\bibitem{dh} G. V. Dunne and T. M. Hall, ``Borel summation of the
derivative expansion and effective actions'', Phys. Rev. {\bf D 60}, 
065002 (1999) [hep-th/9902064].


\bibitem{matt} S. Blau, M. Visser and A. Wipf, ``Analytic Results for
the Effective Action'', Int. J. Mod. Phys. {\bf A6}, 5409 (1991).

\bibitem{ritus} V. I. Ritus, ``Lagrangian of an intense electromagnetic
field and quantum electrodynamics at short distances'', Sov. Phys. JETP
{\bf 42}, 774 (1976).

\bibitem{Leinweber:1993hj}
D.~B.~Leinweber and T.~D.~Cohen,
Phys.\ Rev.\ D {\bf 47}, 2147 (1993)
[hep-lat/9211058].









\end{references}
\end{document}